\documentclass[pre,amsmath,amssymb,twocolumn,floatfix,showpacs]{revtex4}
\usepackage{graphicx} \usepackage{latexsym}
\begin{document}
\title{Breakdown of Lindstedt Expansion for Chaotic Maps}
\author{Guido Gentile}
\affiliation{Dipartimento di Matematica,
Universit\`a di Roma Tre, Roma I-00146, Italy}
\author{Titus S. van Erp}
\affiliation{Laboratoire de Physique / Centre Europ\'een
de Calcul Atomique et Mol\'eculaire, Ecole Normale Sup\'erieure de
Lyon, 46 all\'ee d'Italie, 69364 Lyon Cedex 07, France}

\begin{abstract}
In a previous paper of one of us [Europhys. Lett. 59 (2002), 330--336]
the validity of Greene's method for determining the
critical constant of the standard map (SM)
was questioned on the basis of some numerical findings.  
Here we come back to that analysis and we provide an
interpretation of the numerical results by showing that no
contradiction is found with respect to Greene's method.
We show that the previous results based on the expansion in 
Lindstedt series do correspond to the transition value 
but for a different map: the semi-standard map (SSM).
Moreover, we study the expansion obtained from the SM and SSM
by suppressing the small divisors. 
The first case turns out to be related to Kepler's equation after
a proper transformation of variables.
In both cases we give an analytical solution for the
radius of convergence, that represents the singularity in the complex
plane closest to the origin. Also here, the radius
of convergence of the SM's analogue turns out to be
lower than the one of the SSM. However, despite the absence
of small denominators these two radii are lower than the ones of 
the true maps for golden mean winding numbers.
Finally, the analyticity domain and, in particular,
the critical constant for the two maps without small divisors
are studied analytically and numerically.
The analyticity domain appears to be an perfect circle for the SSM 
analogue, while it is stretched along the real axis for the SM analogue
yielding a critical constant that is larger than its radius of convergence. 
\end{abstract}

\maketitle

\section{introduction}
The Taylor-Chirikov map \cite{Chi79,Gre1979} or standard map (SM) 
is one of the best known nonlinear models showing the onset of chaos
in Hamiltonian systems. It describes with some level of approximation
many physical systems. Among these there are numerous applications to 
plasma-physics, in which field it was originally introduced.
The SM is also exactly related to the time evolution of the 
``kicked rotor" and the equilibrium condition for a chain of masses
superpositioned on a periodic potential. The latter is known as  the
Frenkel-Kontorova (FK) model. This model is of equally importance for 
solid state physics as the SM is for plasma physics. It has, e.g.,
been applied to Josephson junctions arrays, charge density 
waves and surface friction \cite{FloMa96}. 
More importantly, due to the simplicity and, yet, their complex behavior,
these minimalistic models have had an enormous impact for our understanding
in complex phenomena such as nonlinearity, chaos, 
quasi-periodicity, and commensurate-incommensurate transitions.
Although now part of any text-book in nonlinear physics and studied 
extensively over many years, the SM and FK still bear many unsolved
problems. The most intriguing one of these is the sudden transition
from smooth to chaotic orbits in the SM when the coupling parameter
$K$ is increased above a critical value $K_c$. In the FK model
this transition is connected to change from a sliding 
to a pinned state and bears the name Aubry-  or \emph{breaking
analyticity}- transition.

The theoretical framework that characterizes this transition
originates from the Kolmogorov-Arnold-Moser (KAM) theorem \cite{Tabor89}
that deals with the problem of small denominators that can 
occur in any perturbation expansion for quasi-integrable systems.
In fact, the KAM theorem can be used to prove the non-chaotic
behavior of the SM  for very small coupling $K$ and
sufficient irrational winding number $l$.  
Other arguments can then be applied to prove that a chaotic regime 
exists for values $K>K'$ giving an upper bound to $K_c$.
For $l$ equal to the golden mean, there exists an analytical bound by 
Mather $K_c<4/3$~\cite{Math84},
and the computer assisted proof of MacKay and Percival 
$K_c<63/64 \approx 0.9844$~\cite{MaPer85}. 
Moreover, another computer assisted analysis of Jungreis 
excluded the value $K=0.9718$ for possible occurrence
of invariant circles (smooth orbits)~\cite{Jung91}.

There exist several methods to calculate $K_c$ precisely, among which
Greene's method~\cite{Gre1979} has shown to be one of the most effective
giving the estimate $K_c=0.971635$. This method is based
on the assumption that the dissolution of invariant curves can be
associated with the sudden change from stability to instability of
nearly closed orbits. The renormalization method of MacKay~\cite{MacK82} 
is a further refinement of this method and has established the same value
with higher digit precision with respect to the original Greene's result.
Yet, Greene's hypothesis has only been partly proven.
A result by Falcolini and de la Llave~\cite{Faldela1992}
asserts that the critical constants for symplectic maps 
can never be higher than the ones obtained by Greene's method.
Recently, the result has been extended to nontwist maps
by Delshams and de la Llave~\cite{Deldela2000}. Hence, $K_c \leq 0.971635$
for the SM with the golden mean as winding number.

Another way to calculate $K_c$ is through the Lindstedt series expansion.
The smooth orbits in the SM can be described by a continuous
(analytic) periodic function beneath $K_c$. Hence,
below $K_c$ the Fourier spectrum of this function should be finite 
for each component while decaying to zero in the high frequency domain. 
Above $K_c$ some divergence is expected: either infinitely
high frequency components persist, or the amplitude of some
components in the spectrum diverges, or both.
By writing down the Taylor expansion and equating the Taylor orders
in the functional equation satisfied by this conjugation function,
the Fourier-Taylor components can, in principle, be derived from the
ones of lower order. In Ref.~[\onlinecite{ErpFas2002}] an evaluation
of this expansion to high orders suggested a convergence
to a value $K_c \sim 0.97978$, which is higher than Greene's result.
In this article, we come back to this analysis and show that an 
apparent plausible assumption made in Ref.~[\onlinecite{ErpFas2002}]
is falsified beyond Taylor order $n > 200$.
As a result, the Lindstedt expansion does not contradict Greene's result. 
The value $K_c \sim 0.97978$, however, does correspond to
the critical value for a different map, the semi-standard map (SSM).
We come back to this in Sec.~\ref{LinSer}.

Aubry~\cite{Aubry83} proposed another method, that
is probably not very effective for high precision evaluation
in a computer algorithm, but still interesting.  
It is based on an eigenvalue calculation of the dynamical matrix
for the FK chain close to the critical point.
Although this, in principle, requires the diagonalization of an
infinite matrix, one can use the fact that the eigenvector of the 
lowest mode tends to localize~\cite{ErpFas99}.
The instability of the FK chain can than be determined 
in successive approximants by calculating the determinants of 
finite matrices of increasing length.

Finally we mention the use of Pad\'{e} approximants
\footnote{The $(L,M)$-Pad\'e approximation for a function $f(x)$
is given by the ratio of two polynomials, $f(x) \approx P_L(x)/Q_M(x)$,
with $P_L=p_0+p_1 x+\ldots + p_{L} x^{L}$ and
$Q_M=1+q_1 x+ \ldots q_{M} x^{M}$.
Hence, the simple Taylor expansion of order $n$
can be considered as a special case of the Pad\'e series with $L=n$ and
$M=0$.}
to study
numerically the entire analyticity domain. This is a powerful tool
even if is it less precise than other methods
for detecting the critical constant $K_{c}$ and not
completely under control from a rigorous point of view.
It has, for instance, been employed in
Ref.~[\onlinecite{BerCelChiFal1992}] and,
very recently, in Ref.~[\onlinecite{BerFalGen2001}]
where the existence of a natural boundary for the
analyticity domain of the SM has been checked numerically.
Always with the aim of studying the analyticity domain Falcolini and
de la Llave~\cite{Faldela1992b} developed a variant of Greene's method
working for complex values of the parameter $K$ that gives an alternative
to the Pad\'e series approach.

Eventually, these approaches are assumed to converge to 
the same value. However, the proof of this is highly non-trivial.
The ultimate goal, of course,  
would be to gain an analytical expression for $K_c$. 
This is still far beyond our capabilities. 
Inspired to investigate further the influence
of the small denominators in the Lindstedt series expansion,
we introduce two simplified models by setting rigorously all the
denominators equal to one both for the SSM and the SM.
In the latter case, this is a very well 
known model, Kepler's equation~\cite{Wintner1941},
which turns out to have a very similar transition
and can be solved analytically.
The radius of convergence is found to be higher
than the SM and SSM value in case of golden mean winding numbers.

This article is organized as follows:
In Sec.~\ref{SM_SSM} we introduce the SM and SSM. 
In Sec.~\ref{LinSer} we come back to the analysis
of Ref.~[\onlinecite{ErpFas2002}] showing that the Lindstedt expansion
does not violate Greene's method and make
the comparison between the SM and SSM.
In Sec.~\ref{NewMod} we present a new model in which we suppress
the small denominators and give an analytical expression both
for the radius of convergence and the critical constant.
We end with conclusions in Sec.~\ref{Concl}.

\section{The (Semi-) Standard Map}\label{SM_SSM}
The SM and SSM are defined as
\begin{equation}
\left( \begin{array}{c} x_{i+1} \\ x_i \end{array} \right)=
\mathrm{T} \left( \begin{array}{c} x_i \\ x_{i-1}  \end{array} \right)=
\left( \begin{array}{c} 2 x_i + V'(x_i) -x_{i-1} \\
x_i          \end{array} \right) ,
\label{it}
\end{equation}
with 
\begin{eqnarray}
V'(x) &= \frac{K}{2 \pi}\sin(2 \pi x) \quad &\textrm{ for the SM} ,
\nonumber \\
V'(x) &= \frac{K}{4 \pi i}\exp(i 2 \pi x) \quad &\textrm{ for the SSM}. 
\label{V_SM_SSM}
\end{eqnarray}
The resulting sequence $(x_i \textrm{ mod } 1)$ for $i=2, \ldots \infty$
originating from a starting point $(x_1,x_0)$  produces a
discrete trajectory on a two-dimensional torus. 
Such a trajectory for the SM can be 
related to the equilibrium positions of an infinite FK chain where 
particles with harmonic nearest neighbor coupling are placed on a
periodic potential $V(x) = K(2 \pi)^{-2}(1-\cos(2 \pi x))$. 
The SSM has not such a similar counterpart,
but is much simpler in its mathematics. 
By definition, $l \equiv \left \langle x_{i+1}-x_i \right \rangle$
is called the winding number or rotation number of the map. 
For low coupling $K$ and $l$ incommensurate to the periodicity of $V'$, 
there exists a continuous function
$g_{l}(x)$ such that the positions $\{ x_i \}$ can be expressed as
\begin{equation}
x_i=g_{l}(il + \alpha).
\end{equation}
This function is often called the conjugating function or,
in context with the FK model, the modulation or hull function.
The subscript $l$ indicates that the shape of the function
depends on the rotation number. The conjugating function satisfies
the functional equation
\begin{equation}
2g_{l}(x)-g_{l}(x+l)-g_{l}(x-l)=- V'\big(x+g_{l}(x)\big) .
\label{functionalequation}
\end{equation}
For $K$ large enough the function $g_l(x)$ becomes discontinuous. 
For the SM this implies that the orbits become chaotic and for the FK
that the chain of particles gets pinned together
with the appearance of a phonon gap. This transition, in context with
the FK model,
is also called analyticity breaking transition or Aubry transition.

There are several quantities of interest which one can introduce
in order to study the transition from regular to chaotic dynamics.
As the function $g_{l}(x)$ is analytic for $K$ close to the origin
one can consider its series expansion in powers of $K$
\begin{equation}
g_{l}(x) = \sum_{n=1}^{\infty} K^{n} g_{l}^{n}(x) ,
\label{gTay}
\end{equation}
and define the radius of convergence $\rho(l)$ as
\begin{equation}
\rho(l) = \inf_{x\in[0,1]} \Big( \limsup_{n \to \infty}
\left| g_{l}^{n}(x) \right|^{1/n} \Big)^{-1} ,
\label{roc}
\end{equation}
where $\inf_{x\in[0,1]}$ denotes the infimum or greatest lower bound
in the domain $x\in[0,1]$ for the quantity within the brackets 
$(\ldots)^{-1}$ and $\limsup_{n \to \infty}$ is the supremum limit 
giving the highest value for $\left| g_{l}^{k}(x) \right|^{1/k}$
of all $k>n$ in the limit $n\rightarrow \infty$.
Note that the infimum appears in the definition of the radius
of convergence because each invariant curve is filled densely by any 
trajectory lying on it as a result of the incommensurate winding number. 
Hence, existence of the invariant
curve itself requires the latter to be defined for all $x\in[0,1]$.

The critical constant $K_{c}(l)$ is defined as the (positive) real value
$K_{c}(l)$ such that for $K>K_{c}(l)$ the conjugating function
is not analytic any more \footnote{The reason why one usually
does not considers the negative critical constant, that is
the negative value $K_{c}'(l)$ such that for $K<K_{c}'(l)$
there is no more an analytic orbit, is that $K_{c}'(l)=-K_{c}(l)$
for the SM.}. It is believed that the analyticity domain of
the conjugating function has a natural boundary:
this means that $g(K,x)$, which is  the modulation function $g(x)$
at a value $K$, has a set of singularities in terms of $K$ that form
a closed curve around the origin in the complex plane.
Hence, the radius of convergence $\rho$ corresponds to the singularity
closest to the origin, while the critical constant $K_c$ corresponds
to the intersection of this curve with the (positive) real axis.
By definition one has $K_{c}(l) \geq \rho(l)$, so that
by estimating the radius of convergence one finds a lower bound
for the critical constant. Furthermore it is generally accepted
that $K_{c}(\tau)=\rho(\tau)$ for the golden mean
$\tau =(\sqrt{5}-1)/2 \approx 0.618034$ \footnote{The golden mean 
is sometimes in other literature 
defined as the inverse of this value: $(\sqrt{5}+1)/2=\tau^{-1}
\approx 1.618034$.}.
It is also commonly believed (on the basis of numerical simulations
and heuristic arguments) that $K_c(l)$ has the highest value
for the golden mean $\tau$. Moreover, $K_c(l)$ is assumed the same for 
all values $l \in {\mathbb Z}(\tau)$, i.e. the values that can
be written as $l=m \tau+n$ with $m, n$ integer numbers.

So far, the most accurate method to calculate $K_c$ is based on Greene's 
method (also known as residue criterion).
In this method the infinite trajectory $\{ x_i \}$ 
with irrational winding number $l$ is approached by
successive approximants $j$, which are periodic trajectories 
with rational winding numbers $l_j=p_j/q_j$ and
$x_{i+q_j} \textrm{ mod } 1 =x_i$. 
Hence, $q_j$ and $p_j$ are at each level $j$
two integer values giving a better estimate of $l$ for each increment in $j$   
and $l=\lim_{j \rightarrow \infty} l_j$.
These numbers can for instance be obtained using the
continued fraction fraction expansion of $l$. 
For $l=\tau$ this results in the Fibonacci numbers
($\tau \approx F_{j-1}/F_j$ with $F_0=F_1=1$ and $F_j=F_{j-1}  +
F_{j}$ for $j>1$). Conclusively, the Greene's method
tells how to construct the periodic orbits
and to measure their stability by means of the \emph{residue}
that does not tend to zero any more for $K>K_c$.  

Besides only partly proven, Greene's method has also some other
limitations. For instance, this method does not work for other
interesting models, as the SSM and Siegel's problem \cite{Siegel1942},
where the construction of periodic orbits fails
\footnote{One can easily check the non-existence of smooth
periodic orbit by a first orders perturbation theory.}.
The best general alternative is the Lindstedt series expansion.
This method is more generally applicable (it also works for the SSM),
but is less accurate than Greene's method for the SM.

\section{The Lindstedt series expansion}\label{LinSer}
{\bf Standard Map:}
A way to study the transition is by means of the Lindstedt series,
which is the expansion of the function $g_l(x)$ both in Fourier and in
Taylor series
\footnote{Such expansions were originally introduced by Lindstedt
and Newcomb to study problems in celestial mechanics~\cite{Poin1899}.}.
By defining the Fourier transform as
\begin{eqnarray}
g_{l}(x)& =& \sum_{k=-\infty}^{+\infty}
X_k e^{2 \pi i k x} \quad \textrm{with inverse:} \nonumber \\
X_k &=&\int_0^1
\mathrm{d} x \, g_{l} (x) e^{-2 \pi i k x} ,
\label{Fou}
\end{eqnarray}
and expanding
\begin{equation}
X_k=K X_k^1+ K^2 X_k^2 + K^3 X_k^3 + \ldots,
\label{XTay}
\end{equation}
we end up with Fourier-Taylor coefficients $X_k^n$, where
$n$ is the Taylor index  and $k$ is the Fourier index.
Now, using Eq.~(\ref{functionalequation}) 
we can relate the Fourier-Taylor coefficients of order $n$ by
the ones with lower Taylor index by \cite{ErpFas2002}
\begin{eqnarray}
\omega_k^2 X_k^{n} = \frac{i}{4 \pi}
\Big\{ \delta_{1,k}-\delta_{-1,k} \Big\} \delta_{1,n} +
\frac{i}{4 \pi} \sum_{m=1}^{\infty} \frac{(i 2 \pi)^m}{m!} \times 
\nonumber \\
\sum_{  n_1+n_2 +\ldots+ n_m=n-1}  \Big\{
\sum_{ k_1+ k_2+ \ldots +k_m=k-1 }
 X_{k_1}^{n_1} X_{k_2}^{n_2}\ldots X_{k_m}^{n_m} \nonumber \\ 
- (-1)^m 
\sum_{ k_1+ k_2+ \ldots +k_m=k+1 }
X_{k_1}^{n_1} X_{k_2}^{n_2}\ldots X_{k_m}^{n_m}
  \Big\} ,
\label{TayFou}
\end{eqnarray}
with 
\begin{eqnarray}
\omega_k^2 &=& \frac{1}{X_k} \int_0^1
\mathrm{d} x \, \big( 2 g_{l}(x)-g_{l}(x+l)-g_{l}(x-l) \big) \,
e^{-2 \pi i k x} \nonumber \\
 &=& 2\big(1 - \cos(2 \pi k l) \big)=\big( 2 \sin(\pi k l) \big)^2 ,
\label{omega}
\end{eqnarray}
and where $\sum_{n_1+n_2+\ldots+n_m=n-1}$ implies
a summation of all possible integers $n_1, n_2, \ldots, n_m$ with
the constraint that $\sum_{i=1}^m n_i=n-1$.
There are ways to reduce the number of summations in Eq.~(\ref{TayFou}).
One possible way was proposed in Ref.~[\onlinecite{ErpFas2002}]
to construct an extended matrix $P(n,k,m)$ defined as
\begin{eqnarray}
P(n,k,m)=\frac{(2 \pi i)^m}{m!}  
\sum_{n_1 + n_2 +\ldots+ n_m=n}  
\nonumber \\
\Big[
\sum_{k_1+ k_2+ \ldots+ k _m=k} 
X_{k_1}^{n_1}X_{k_2}^{n_2}\ldots X_{k_m}^{n_m}  \Big] .
\label{defP}
\end{eqnarray}
One can show that $P(n,k,m)=0$ if $|k|>n$ or $m>n$. 
This gives rise to following recursive relations~\cite{ErpFas2002}:
\begin{eqnarray}
P(1,\pm 1,1)&=&\frac{\mp 1}{2 \omega_1^2} , \label{recP}  \\
P(n,k,1)&=& -\frac{1}{2} \omega_k^{-2} \sum_{m=1}^{n-1} \Big[ P(n-1,k-1,m)
\nonumber \\
&-&(-1)^m P(n-1,k+1,m) \Big] ,
 \nonumber \\
P(n,k,m)&=&\frac{1}{m} \sum_{n'=1}^{n-m+1} \, 
\sum_{k'=\max\{-n',k-n+n' \} }^{\min\{n',k+n-n'\}}
 \nonumber \\
&&P(n',k',1) P(n-n',k-k',m-1), \, m >1 , \nonumber 
\end{eqnarray}
from which we can distract the Fourier-Taylor coefficients by
\begin{equation}
X_k^n=\frac{P(n,k,1)}{2 \pi i} .
\label{XP}
\end{equation}
The components of the $P$ matrix are all real
and obey the symmetry relation $P(n,k,m)=(-1)^m P(n,-k,m)$.
Moreover, besides being zero for $|k|>n$ and $m>n$,
the matrix $P(n,k,m)$ has zero values whenever $k+n$ is odd. Hence,
$k=n, n-2, \ldots, -n$ are the only nonzero elements of $P$.

\begin{figure}[!h]
\includegraphics[width=8cm,keepaspectratio]{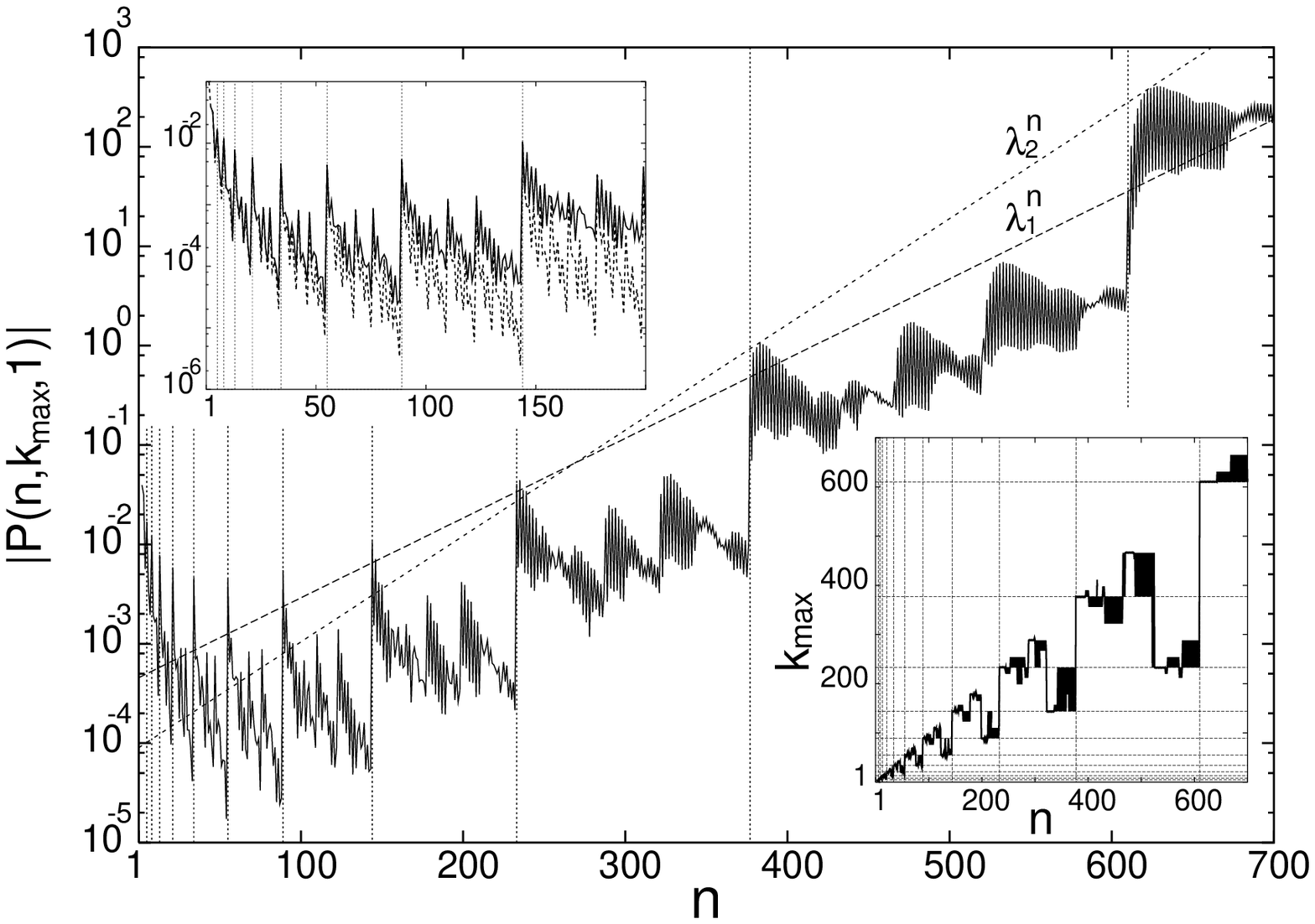}
\caption{$|P(n,k_{\rm max},1)|$ as a function of $n$.
This is defined as the maximum value of $|P(n,k,1)|$ of all $k$.
Hence, $k_{\rm max}$ is defined as the $k$ value where 
$|P(n,k,1)|$ has this maximum.
 The inset in the lower corner shows $k_{\textrm{max}}$ as
function of $n$.
The inset in the left upper corner is an enlargement of the first 
200 terms together with $|P(n,n,1)|$ (dashed line).
From these figures, one can clearly detect sudden boosts
in the function $|P(n,k_{\rm max},1)|$ where 
$k_{\rm max}=n$ at the Fibonacci values  
(dashed vertical lines). However, whereas for $n<200$ 
the character is sharply peaked at these values, its behavior changes
for higher orders. Still $k_{\rm max}=n$ for $n$ a Fibonacci number,
but the intersecting line described by 
$\alpha_1\lambda_1^n$ does no longer dominate
the complete evolution of all the $|P(n,k,1)|$ terms.
$\lambda_1=1.0186$ is determined by the line through $(n,k)=(F_{13},F_{13})=
(377,377)$ and $(n,k)=(F_{14},F_{14})= (610,610)$.
$\lambda_2=1.0248$ is set by the line through $(n,k)=
(383,377)$ and $(n,k)= (622,610)$ where $|P(n,k_{\rm max}(n),1)|^{1/n}$ 
shows local maxima in $n$. The inversed values, $\lambda_1^{-1}
\approx 0.9817$ and $\lambda_2^{-1} \approx 0.9758$ are assumed
to converge for higher $n$ to $K_c$ for the SSM and SM, respectively.}
\label{fig1}
\end{figure}

The evaluation of Eqs.~(\ref{recP}) is very efficient
to evaluate $X_k^n$  obtaining a Taylor order of approximately $n=200$.
To go beyond this limit, sufficient computer power and time is needed
as both the computation time as the number of nonzero  
matrix elements increase with $\sim n^3$. 
Hence, memory can become a severe problem as the number of components
that have to be stored can easily go beyond the 
maximum allowed allocation limit
of the computing system. In this work, 
we finally reached the level $n=700$ (see Fig.~\ref{fig1}) and we believe that
going beyond this order is not very profitable for obtaining
a more accurate evaluation of $K_c$. We come back to these results
after addressing the small denominator problem that arises from 
Eq.~(\ref{recP}). 

From Eq.~(\ref{omega}) and the second line in Eq.~(\ref{recP}) one sees 
that even for irrational values of $l$, the terms $\omega_k^{-2}$
can become arbitrarily high for some $k$. This effect is a
typical example of the `small denominator problem' that can
strongly prevent the convergence of any perturbative series.
In fact, it requires a stronger condition than irrationality
such as a Diophantine condition~\cite{Tabor89}.
Among all the irrational numbers, the golden mean $\tau$ suffers the
least from the small denominator problem and has therefore
the highest convergence radius $K_c$.
The golden mean is relatively difficult to approximate by
rational numbers as those arising, for instance, from
the continued fraction expansion (one can say that
it is the `most irrational number'). As addressed above,
the best approximants for the golden mean are given by the Fibonacci numbers:
$\tau=F_j/F_{j+1}$. Therefore, the small denominators
will be most dominant for $k=F_j$. 
From the exact relation $F_{j-1} -F_{j} \tau = (-\tau)^{j+1}$,
one can show that for $j\to\infty$ 
\begin{equation}
\frac{1}{\omega_{F_j}^2} \approx \frac{1}{4 \pi^2} 
\Big(\frac{1}{\tau^2}\Big)^{j+1} \sim (2.618)^{j+1} .
\label{divom}
\end{equation}
However, we would like to stress 
that the small denominator problem is not the only mechanism 
causing the breakdown of the perturbative approach.
This becomes evident in Sec.~\ref{NewMod} where we introduce the model that
arises when we rigorously set $\omega_{k}^{-2}=1$ for all $k$
in the series of Eq.~(\ref{recP}). Clearly, this simplified expansion
can not be affected by the small denominators. However, it still has a 
radius of convergence and a critical constant,
as shown by the analytical solution.
As the radius of convergence $\rho$ of this simplified model 
is found to be lower than $\rho(\tau)=K_c(\tau)$ for the SM, it proves
that the golden mean winding numbers $l \in \mathbb{Z}(\tau)$
are remarkably resistant to the problem of small denominators.
The full analysis of this model is given in Sec.~\ref{NewMod}.
 
Coming back to the results of Fig.~\ref{fig1}, 
we see that indeed the evolution of $P(n,k,1)$ makes
sudden jumps at the Fibonacci numbers as expected from Eq.~(\ref{divom}).
Besides, deceptively the values $n=k=F_n$ seems to determine the
whole power law behavior of $X_k^n$, which is true till Taylor order
$n\sim 200$. This assumption made in Ref.~[\onlinecite{ErpFas2002}]
allows for a further simplification  of Eq.~(\ref{recP}) by defining
the reduced matrix $Q(n,m)\equiv P(n,n,m)$ obeying the relations 
\begin{eqnarray}
Q(1,1)&=&\frac{- 1}{2 \omega_1^2} , \nonumber \\
Q(n,1)&=& -\frac{1}{2} \omega_n^{-2} \sum_{m=1}^{n-1}  Q(n-1,m) ,
\label{recSSM} \\
Q(n,m)&=&\frac{1}{m} \sum_{n'=1}^{n-m+1} 
Q(n',1) Q(n-n',m-1) . \nonumber
\end{eqnarray}
This set of equations make high order ($\sim n=F_{20}=10946$) 
evaluations accessible for computer calculations. At this order
the value of the radius of convergence seems to 
stabilize near $\rho = 0.97978$ , but this is still
higher than the one obtained by Greene's method.
As consequence, validity of Greene's method
was questioned in the quoted paper~\cite{ErpFas2002}.

The more elaborated calculations in this work show
that the assumption made in Ref.~[\onlinecite{ErpFas2002}] was actually 
wrong as shown by the high order behavior in Fig.~\ref{fig1}.  
Still, we find that $k=n$ gives the maximum for $|P(n,k,1)|$ 
whenever $n$ is a Fibonacci number.
However, the character of the evolution changes from being peaked
to more smooth oscillations. Clearly, the line connecting
$|P(F_j,F_j,1)|$  does no longer dominate 
the increment of $P$-matrix elements for $n>200$.
As $|P(n,k_{\rm max}(n),1)|^{1/n}$ shows local
maxima for $(n,k_{\rm max})=(383,377)$ and 
$(n,k_{\rm max})=(622,610)$ just after $F_{13}$ and $F_{14}$, we
fitted the line $\alpha_2 \lambda_2^n$ through these points.
From this fit, $\lambda_2=1.0248$, the estimate
for $K_c \approx \lambda_2^{-1}=0.9758$ is obtained.
Although still higher than Greene's value,
it is already considerably lower than $\rho=0.97978$ obtained from
Eqs.~(\ref{recSSM}) for $n=F_{20}=10946$~\cite{ErpFas2002}.
Note that the latter approach of Ref.~\cite{ErpFas2002} for this lower 
approximant $n=F_{14}=610$, as obtained from the line $\lambda_1^n$
(See Fig.~\ref{fig1}), would result in $\rho \approx 1/\lambda_1=0.9817$,
still approximately $0.002$ higher than the 
nearly converged value of $\rho=0.97978$.
Hence, a decay of $0.004$ from $0.9758$ at $n=622$ to $0.9716$ 
at $n\rightarrow \infty$ is not unlikely.  As a consequence, contrary
to the results of the restricted series~(\ref{recSSM}), 
there is no proof that the full Lindstedt series~(\ref{recP})
violates Greene's hypothesis. This also shows that further
simplification of Eqs.~(\ref{recP}) is not easily obtained and that an 
accurate evaluation of $K_c$ based on the Lindstedt perturbation is
severely demanding.

\vskip.3truecm
{\bf Semi-Standard Map:}
When evaluating Eq.~(\ref{it}) for the SSM~(\ref{V_SM_SSM}),
the factors $\delta_{-1,k}$ and $-(-1)^m\sum \ldots$ are
not present in Eq.~(\ref{TayFou}). It is then straightforward to show that 
the $P$-matrix~(\ref{defP}) has only non-zero elements for $n=k$. Hence, 
the assumption made in Ref.~\cite{ErpFas2002} that gave rise to
Eqs.~(\ref{recSSM}) does not corresponds to the critical constant
of the SM, but still gives the correct value for SSM. A result by
Davie~\cite{Dav1995} shows that, for maps including the ones we are
considering, the radius of convergence (\ref{roc}) is equal to
\begin{equation}
\rho(l) = \Big( \limsup_{n \to \infty} \max_{|k|\le n}
\left| X^{n}_{k} \right|^{1/n} \Big)^{-1} .
\label{rocdavie}
\end{equation}
Therefore, the radius of convergence for the SM cannot be larger
than the radius of convergence of the SSM, but of course
it implies only a lower bound on the critical constant.
Numerically by using Pad\'e approximants
in Ref. \cite{BerFalGen2001} it has been found that
for certain values of the rotation number $l$, the radius of convergence
of the SM is strictly smaller than the radius for the SSM.
For the golden mean it is hard to improve upon simple power series
using Pad\'e, but for other numbers closer to resonant values
it is possible and the phenomenon becomes much more evident. 

The fact that the critical constant for these numbers
is lower for the SM than for the SSM implies,
by (\ref{rocdavie}), that dominant contributions arise
from terms with Taylor orders $n$ for which 
$|X_{k_{\rm max}}^{n}|> |X_{n}^{n}|$. 
This is exactly what emerges from the numerics as 
noted above and shown in Fig.~\ref{fig1} for $n>200$.
Clearly, this is not be the case for the SSM where one can limit to $k=n$.
Hence, in Ref.~[\onlinecite{ErpFas2002}] $K_c(\tau)=\rho(\tau)$
was actually determined
for the SSM to be $0.97977$ at Taylor order $n=F_{20}=10946$.
In the calculations of this work, we went to order $n=F_{24}=75025$, 
that gave the value $0.97937$.
As the root criterion saturates very slowly, 
the numerical results provide essentially only an
(accurate) upper bound for the radius of convergence.

To conclude this section, we found that,
also for the golden mean, the radius
of convergence for the SM is strictly less than the
radius of convergence for the SSM.
Therefore, as a general comment we can remark that for the SM
the presence of all harmonics in the Fourier expansion
of the coefficients $g^{n}_{l}(x)$ has a double effect. 
On  one hand the radius of convergence becomes smaller
with respect to that of the SSM. On the other hand,
the critical constant $K_c(l)$ can be larger than its
radius of convergence $\rho(l)$ for $l \notin \mathbb{Z}(\tau)$.
For the golden mean the two values are equal
as emerges numerically~\cite{Faldela1992b}, but for other values
they can be appreciably different.
One can imagine that the first phenomenon is due to the
presence of contributions $X^{n}_{k}$ larger than
$X^{n}_{n}$, while the second one is a consequence of deep
cancellations between the harmonics of given perturbative order.
These two effects are, in general, much 
more dominant for rotation numbers $l$ close to rational
values (see for instance Refs.~\cite{BerFalGen2001,BerGen1999,BerGen2001}).

\section{Putting small denominators to unity}\label{NewMod}
\subsection{Introduction of the simplified maps}
\begin{figure}[!h]
\includegraphics[width=8cm,keepaspectratio]{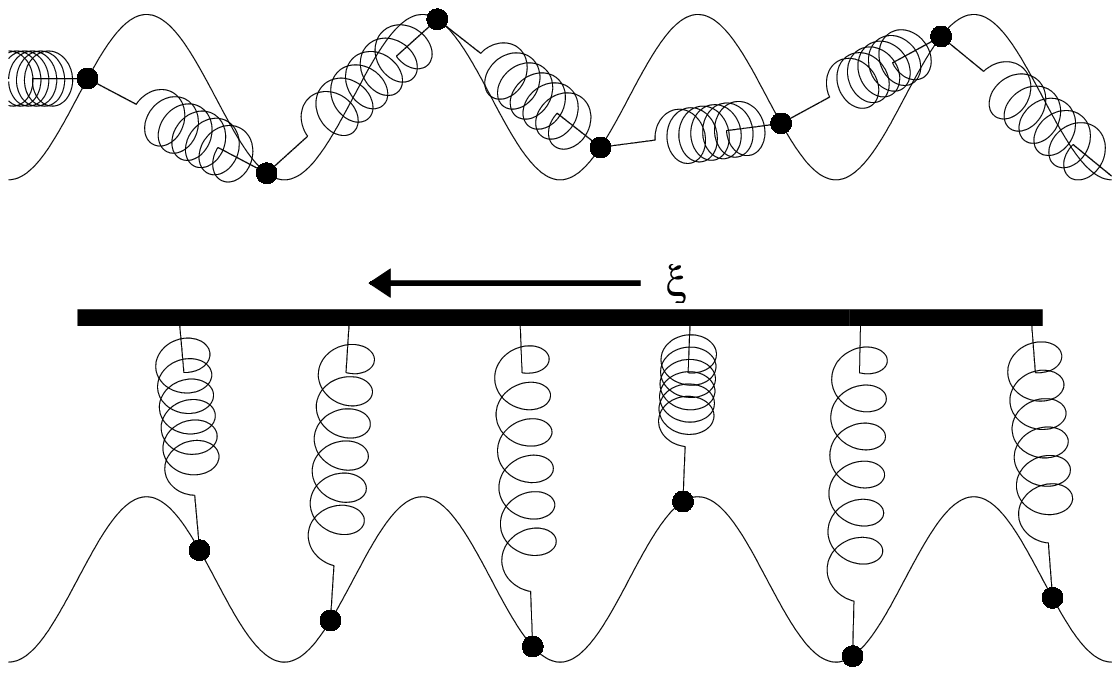}
\caption{Illustration of the FK model (top) and the system 
that obeys Eq.~(\ref{funceq2}) (bottom). The latter corresponds
also to the FKT model without neighbor interaction. All particle
are connected to the upper rod whose position is given by 
$\xi$. A sliding mode may exist when $\xi$ can be varied.
 }
\label{fig2}
\end{figure}

An interesting study  appears if we set rigorously
all possible small denominators equal to unity:
$\omega_k=1$ for all $k$ in Eqs.~(\ref{recP}) and (\ref{recSSM}). 
Although the inspiration of this model was simply 
to study of the perturbation expansion when the 
small denominators have no effect, we can retrace from this series
back to a functional relation as the one in Eq.~(\ref{functionalequation}). 
It can be shown that these simplified series
would correspond to the functional relation for a function $h(x)$
\begin{eqnarray}
h(x)+\frac{K}{4 \pi i } \exp\big(i 2 \pi \big(x+h(x) \big) \big)&=&0
\textrm{ for the SSM and} \nonumber \\
h(x)+\frac{K}{4 \pi} \sin \big( 2 \pi \big(x+h(x)\big)&=&0 
\textrm{ for the SM}.\label{funceq2}
\end{eqnarray}
Hence, the divergence of the simplified series corresponds
to the point in $K$ where these functional forms~(\ref{funceq2}) 
have no analytical solution any more.
A logical next step would be to relate the equality~(\ref{funceq2})
to the iteration of a map similar to Eq.~(\ref{it}).
As the relation~(\ref{funceq2}) does no longer contain
the arguments $x\pm l$ this is not so evident.
However,
one can relate the $h(x)$ function to the 
hull function of a FK-type system.
It can be shown that 
this corresponds to an one-dimensional 
Einstein solid that is interacting with an external incommensurate
potential. Due to the lack of neighbor interaction, 
which makes each particle independent, it is highly unusual 
to  describe for such a system the equilibrium coordinates
by a collective hull function.
Still, there are no restriction not to do so and one can even give such
a function a physical meaning. As known from the FK model,
the continuous shape of the hull function
is directly associated with the existence of a sliding mode
where the FK chain can slide over the periodic potential without
cost of energy \cite{FloMa96,ErpFas99}. 
In this case, the complete phonon spectrum is 
giving by sum of oscillations of the individual particles that are
not zero in general. The sliding mode appears 
when we add an extra degree of freedom to the system as shown in
Fig.~\ref{fig2}.

Here, all particles have no interaction to their neighbors,
but are connected to an upper rod. When the rod has an infinite mass
compared to the particle masses, the system is basically an Einstein solid.
However, if we assume that the position of the rod may vary according to a 
coordinate $\xi$ an extra phonon mode exist that is zero for $K<K_c$
in this system. Hence, the breakdown of the
Lindstedt expansion~(\ref{recP}) with $\omega_k=1$ for all $k$
can be also related to a real physical sliding-pinning transition.
The model as illustrated in Fig.~\ref{fig2} 
is also equal to a special case of the Frenkel-Kontorova-Tomlinson 
(FKT) model where each particle is connected to a rod by a spring
with spring constant $c_r$ and additionally to its 
neighbors with a coupling $c_n$~\footnote{The simple 
Tomlinson model~\cite{Tom29} is not the system in
Fig.~\ref{fig2} as this usually consists of only one single oscillator 
(particle). This ancient model is now often applied to simulate
the ``stick-slip" motion of an atomic force microscope(AFM)-tip over
a sample surface.}. The FKT model has been proposed
to study more realisticly  the frictional behavior between atomic 
surfaces~\cite{WeiEl96,WeiEl97,GyaTho97}. 
The model that is described by Eq.~(\ref{funceq2}) simply corresponds
to the FKT system with $c_r=1$ and $c_n=0$.
The nice thing is that the perturbation series of Eqs.~(\ref{funceq2})
can be solved exactly. To show this, we will start with the more simple 
SSM case.
\subsection{Radii of convergence for simplified maps}
{\bf Simplified SSM:}
It is convenient to use following normalization
\begin{equation}
R(n,m)=n! (-2)^n Q(n,m) ,
\label{defR}
\end{equation}
with matrix entries that are integer and positive
and with $R(n,n)=1$. 
From Eq.~(\ref{recP}) and (\ref{defR}) with $\omega_n^{-2}=1$
we derive
\begin{eqnarray}
R(n,1) &=& n \sum_{m=1}^{n-1} R(n-1,m) , \label{recR}\\
R(n,m) &=& \frac{1}{m} \sum_{n'=1}^{n-m+1} \binom{n}{n'} 
R(n',1) R(n-n',m-1) ,\nonumber
\end{eqnarray}
with $\binom{n}{n'}$ the binomial coefficient: $\equiv
\frac{n!}{n'! (n-n')!}$.
Note that the recursive relations in Eq. (\ref{recR}) for $m>1$
coincide with those satisfied by the Stirling numbers
of first and second kind, $S^{(m)}_{n}$ 
and $\mathfrak{S}^{(m)}_{n}$ respectively 
\footnote{The Stirling numbers of the first kind
$S^{(m)}_{n}$ are defined by the requirement
that $(-1)^{n-m}S^{(n)}_{m}$ is the number of permutations of
$n$ symbols which have exactly $m$ cycles.
The Stirling numbers of the second kind
$\mathfrak{S}^{(m)}_{n}$ are equal to the way of partioning a set of
$n$ elements into $m$ non-empty subsets. See
\cite{AbrSte1972}, p. 824, \S 24.1.3 and \S 24.1.4, with $r=m-1$.}.
Of course what is different is the relation for $m=1$.

From relations~(\ref{recR}) with $R(1,1)=1$ the following exact 
equality can be proven
\begin{eqnarray}
R(n,m)&=& 
n^{n-m}\binom{n-1}{m-1}.
\label{relbinomial}
\end{eqnarray}
The proof of this equation is given in the appendix in two ways.
In App.~\ref{secindR} we derive this proof using the argument of induction.
In App.~\ref{sectreeR} we give a proof based on the tree formalism that 
was firstly introduced in Ref.~[\onlinecite{Gal1994}].
The first proof is straightforward, but elongated.
The second is short, but less self-contained as it requires some knowledge
of the previous publications about the tree formalism.  
The latter is, however, most practical for the more complicated 
proof of relation  
for the SM in series of Eq.~(\ref{funceq2}) with $\omega_k^{-2}=1$.
Now, from Eq. (\ref{defR}) and (\ref{relbinomial}) we deduce that
\begin{equation}
Q(n,1) = (-1)^n \frac{n^{n-1}}{2^n n!} ,
\label{relQ}
\end{equation}
which is assumed to increases as a power law $\sim a \lambda^n$
giving the radius of convergence as $\rho =1/\lambda$. Hence,
\begin{eqnarray}
\ln |Q(n,1)| &=& (n-1) \ln(n) - n \ln(2)-\ln(n!) .
\end{eqnarray} 
Then using~\cite{Wells86}
\begin{eqnarray}
\ln(n!) &=& \Big( n + \frac{1}{2} \Big) \ln n -n+\frac{1}{2}\ln(2\pi),
\label{Stirling}
\end{eqnarray}
that is an refinement of the well known Stirling's 
formula $\ln n!\approx n \ln n-n$, we get
\begin{eqnarray}
\ln |Q(n,1)| &=& 
n ( 1-\ln(2) ) - \frac{3}{2}\ln(n)-\frac{1}{2} \ln(2\pi)  \nonumber \\
\Rightarrow |Q(n,1)| &\sim & \frac{1}{\sqrt{2 \pi n^3}} \left(
\frac{1}{2} e \right)^n ,
\label{Qexp}
\end{eqnarray}
yielding a radius of convergence $\rho=2/e \approx 0.735759$.
This value is less than the SSM value or SM value.
This is a bit contra-intuitive, as one would expect
that the possible occurrence of small denominators would give a
lower $\rho$. Apparently, this does not happen for the golden mean.
This can be understood by following reasoning.
Although the small denominator factors $\omega_k^{-2}$ can become
arbitrary large for some $k$ giving a boost to 
the series~(\ref{recP}) and~(\ref{recSSM}), at most values of $k$ they
will be considerable smaller than 1 resulting in an 
opposite effect. Hence, for the golden mean as winding number
the latter effect seems to be more dominant yielding an
even higher value for $\rho$ than the case where
all $\omega_k^{-2}$ terms are equal to 1. 

{\bf Simplified SM:}
The simplified SM considered in Eq.~(\ref{funceq2})
is well known in celestial mechanics~\cite{Wintner1941}
after applying the following variable transformation.
Write $K/2=-{\rm e}$ and $2 \pi x={\rm M}$
where ${\rm e}$ is the eccentricity and ${\rm M}$
is the mean anomaly. Then the eccentric anomaly ${\rm E}=2 \pi (x+h(x))$ 
is related to ${\rm M}$ through Kepler's equation
${\rm M}={\rm E}-{\rm e} \sin {\rm E}$,
which is exactly the second equation in Eq.~(\ref{funceq2}).

The recursive relations~(\ref{recP}) with $\omega_k^{-2}=1$ 
have also an exact solution that we write here:
\begin{eqnarray}
P(n,k,1) & = &  
 \frac{(-1)^{n+(n-k)/2}}{2^{n}}
\frac{k^{n-1}}{((n-k)/2)!((n+k)/2)!} ,  \nonumber \\
&& \textrm{for} |k|<n \textrm{ and if } k+n \textrm{ is even}\nonumber \\
&=& 0 \textrm{ otherwise} ,
\label{resultPnk1}
\end{eqnarray}
which can be obtained by the Lagrange inversion
theorem~\cite{WW1921,Wintner1941}.
We present a new derivation of this relation
based on the tree formalism in the App.~\ref{treeproof2}

Then, by using Eq.~(\ref{rocdavie}) 
we see that we have to compute the maximum over $k$ of $|P(n,k,1)|$.
By assuming that the maximum is reached for some $k$ which is not too
close to $n$ (an assumption that we shall verify
a posteriori \footnote{It is immediate to realize that
the maximum is reached for some $k\ge 1/2$.}), we can approximate
the factorial appearing in Eq.~(\ref{resultPnk1})
with Stirling's formula~(\ref{Stirling}). This gives rise to
\begin{eqnarray}
\left| P(n,k,1) \right|   
& \sim & \frac{1}{2^{n}}
\frac{k^{n-1} e^{n}}{ (\frac{1}{2}(n-k))^{\frac{1}{2}(n-k)} 
(\frac{1}{2}(n+k))^{\frac{1}{2}(n+k)}} \nonumber \\
&\times&
\frac{1}{
(\frac{1}{4}(n^{2}-k^{2}))^{\frac{1}{2}} } \label{Pnk1} \\
& \sim & \frac{2 e^{n}}{n^{2}} \frac{1}{\sigma(1-\sigma^{2})^{\frac{1}{2}}}
\nonumber \\ 
&\times&
\left( \frac{\sigma}{(1-\sigma )^{\frac{1}{2}(1-\sigma)} (1+\sigma)^{
\frac{1}{2}(1+\sigma)} } \right)^{n} , \nonumber 
\label{apprcoeff}
\end{eqnarray}
where we have defined $\sigma=k/n \in [-1,1]$.
Hence, we have to compute the maximum of the function
\begin{eqnarray}
 E(\sigma) &=&
\sigma \exp \left[ - \frac{1-\sigma}{2}\ln (1-\sigma) -\
\frac{1+\sigma}{2}\ln (1+\sigma) \right]. \nonumber \\
\label{function}
\end{eqnarray}
By taking the derivative $\frac{\partial E(\sigma)}{\partial \sigma}=0$,
we find that the maximum is reached at a value $\sigma_{\rm max}$
that satisfies following relation:
\begin{eqnarray}
2+\sigma_{\rm max}\ln(1-\sigma_{\rm max})-
\sigma_{\rm max}\ln(1+\sigma_{\rm max})=0 ,
\label{relxmax}
\end{eqnarray}
yielding $\sigma_{\rm max} \approx 0.833557$. 
Hence,
$k_{\rm max} = \sigma_{\rm max} n \approx 0.833557 n$.
Using Eq.~(\ref{relxmax}), $E( \sigma_{\rm max})$ simplifies to
$E( \sigma_{\rm max})=\frac{1}{e}\frac{\sigma_{\rm max}}{\sqrt{1-
\sigma_{\rm max}^2}}$. Inserting this relation into Eq.~(\ref{Pnk1}) gives 
\begin{equation}
|P(n,k_{\rm max},1)| \sim
\frac{2 }{n^{2}\sigma_{\rm max}^{2}} \lambda^{n+1} ,
\end{equation}
with $\lambda=\frac{\sigma_{\rm max}}{\sqrt{1-\sigma_{\rm max}^2}}$.
This yields a radius of convergence $\rho=\lambda^{-1} \approx 0.662743$,
which is known as the Laplace limit~\cite{Finch2003}.
This value is again smaller than the radius
of convergence of the true SM (recall that for the golden mean the
radius of convergence $\rho(\tau)$ equals $K_{c}(\tau)$).
Moreover, similar to the true maps, this SM-analogue
transition value $\rho$ is lower than the one of the SSM.

\subsection{The critical constant and the analyticity domain}
The argument above gives only information about
the location of the singularities closest to the origin.
The solution of the functional equations~(\ref{funceq2})
could still exists for real values of $K$ larger than the radius $\rho$.
This is the situation $K>\rho$ where $h(K,x)$ is still analytical,
but at which the power series~(\ref{gTay}) and~(\ref{XTay}) are no
more defined. This is typical for a summation that consists of both
positive as negative terms. In particular, there could be no
singularity at all on the real axis so that an analytical form
of $h(K,x)$ could still exists for $K\rightarrow \infty$. 

To analyze the extent of the analyticity domain and the critical constants,
we need to `evaluate' the summations of Eqs.~(\ref{XTay}) and~(\ref{Fou}). 
This means that we need to find functional form $h(K,x)$ that corresponds
to the power series, but, contrary to the summation itself, can still be 
perfectly defined for $|K|>\rho$.

In the following analysis, we will show that the analyticity domains
for the simplified maps are, like the for true maps,
also constrained by a closed boundary.
More precisely, we find that for fixed $x$ there are only a few
singularities, but the union over all $x\in[0,1]$
of such singularities reconstruct a closed curve surrounding the origin.
Hence, outside this natural boundary there is no function $h(K,x)$
that can be obtained by an analytic continuation of
the power series around $K=0$. Although very unlikely,
this does not completely exclude the existence
of a very different function, say $\hat{h}(K,x)$, that is defined
outside this domain and obeys Eq.~(\ref{funceq2}) and may even persist
for $K\rightarrow \infty$. Recurrence phenomena of this kind are known
to occur for certain maps~\cite{Wilbrink1987}, but, for instance, 
this is not the case of the SM.

{\bf Simplified SSM:}
As for this model, $X_k(K)=X_k^k K^n$ we can directly
writing down the summation of Eq.~(\ref{gTay}) for $h(K,x)$ function:
\begin{eqnarray}
h(K,x)=\sum_{k=1}^{\infty} X_k^k K^k e^{2 \pi i kx} .
\end{eqnarray}
Inserting the expression~(\ref{relQ}) gives
\begin{eqnarray}
h(K,x)&=&\frac{1}{2 \pi i} \sum_{k=1}^{\infty} (-1)^k 
\frac{k^{k-1}}{2^k k!} K^k e^{2 \pi i kx}=\nonumber \\ 
&=& \frac{1}{2 \pi i} \sum_{k=1}^{\infty}
|Q(k,1)| K^k e^{\pi i k(2x+1)}
\label{sumhssm}
\end{eqnarray}

To find the full analyticity domain of $h(K,x)$,
one basically has to fix a certain value for $x$, say $x=x'$,
and search for the singularities in $K$ of the function $h(K,x')$ by
e.g. using the Pad\'e approximants method.
Then, one has to repeat, in principle, this procedure for
all possible values of $x$ and collect the set of all singularities
to construct the full analyticity domain. 
Finally, the radius of convergence $\rho$
is then the complex singularity closest to the origin,
while $K_c$ is the smallest (positive) real singularity, if any.

Vice versa, we could also fix the argument $\phi$ of the complex value 
$K$, such that $K=|K| e^{i\phi}$. The summation~(\ref{sumhssm}) 
will then be maximized for $x=-(\frac{\phi}{2 \pi}+\frac{1}{2})$,
where each term in the sum turns into a positive value.
For these values of $K$ and $x$, using the inclusion argument
and the root criterion on the delimiting series,
one can show that the radius of convergence is given by $2/e$.
Hence, for each $x$ there is one singularity at $K=-\frac{2}{e}
e^{-i 2\pi x}$, and the complete set over all $x$ forms a natural
boundary that is a circle around the origin.

Note that this is very different from the true maps. Although not proven,
numerical studies (for instance Ref.~\cite{BerFalGen2001}
and references quoted therein) suggest that for the SM and SSM the
function $g(K,x)$ has for each value of $x$, independent to its value,
an infinite set of singularities forming the same (for each $x$) 
natural boundary. Numerical analysis~\cite{BerFalGen2001} 
shows that the natural boundary of the SSM is 
a circle, just as in this simplified model. 
This property appears to be true irrespective to the choice of $l$
as long as it fulfills Diophantine condition~\cite{Tabor89}. 
For the SM with golden mean as winding number this
curve resembles close to a circle, but not very smooth and slightly
elongated (about 1\%) along the imaginary axis~\cite{Faldela1992b}.

{\bf Simplified SM:}
Taking the power series~(\ref{XTay}) for $X_k$ for the simplified SM 
using Eq.~(\ref{resultPnk1}) we have
\begin{eqnarray}
X_k(K)&=&\sum_{n=1}^{\infty} \frac{1}{2\pi i} P(n,k,1) K^n\nonumber \\
&=& \frac{1}{2\pi i} \sum_{n=|k|,|k|+2, \dots}^{\infty}
K^n  \frac{(-1)^{n+(n-k)/2}}{2^{n}}\nonumber \\
&\times&
\frac{k^{n-1}}{((n-k)/2)!((n+k)/2)!} . \nonumber \\
\end{eqnarray}
Changing variables to $j=(n-|k|)/2$ gives:
\begin{eqnarray}
X_k(K)&=&  
\frac{(-1)^{k}}{2\pi i k} \sum_{j=0}^{\infty}
\frac{(-1)^j}{2^{2j+|k|}j!(j+|k|)!}
(|k|K)^{2j+|k|}   \nonumber \\
&=& \frac{(-1)^{k}}{2 \pi i k} J_{|k|}(K|k|) ,
\label{XtoB}
\end{eqnarray}
with $J_v(z)$ the Bessel function of the first
kind~\cite{GraRyz2000,Watson1944} defined (for integers $v$) as
\begin{eqnarray}
J_v(z)\equiv \sum_{j=0}^{\infty}
\frac{(-1)^j}{2^{2j+v} j!(j+v)!} z^{2j+v} .
\label{defBesselJ}
\end{eqnarray}
As these Bessel functions $J_v(z)$ have no singularities in $z$,
neither has $X_k(K)$ in $K$. Therefore,  the Fourier components do not
give direct information about $K_c$. On the other hand, one can conclude
from Eqs.~(\ref{XtoB})  and~(\ref{defBesselJ})
that $|X_k(K)|$ is maximized for pure imaginary $K$,
so that the radius of convergence is lying on the imaginary axis
on a distance $\rho$ from the origin.
Here, the individual terms in Eq.~(\ref{defBesselJ}) can not cancel
as $(-1)^j$ is then neutralized by $x^{2j}\sim K^{2j}=(-1)^j|K|^{2j}$.
Hence, we expect to obtain the singularity equal to the radius 
of convergence for $K$ along the imaginary axis.

We can now try to evaluate the Fourier series~(\ref{Fou}) for $h(x)$:
\begin{eqnarray}
h(K,x)&=&\sum_{k=-\infty}^{+\infty} \frac{(-1)^{k}}{2 \pi i k}
J_{|k|}(|k| K) \exp(2 \pi i k x)\nonumber \\
&=& \sum_{k=1}^{\infty} \frac{(-1)^{k}}{\pi k}  J_{k}(k K)
\sin(2 \pi k x) .
\end{eqnarray}
Further simplification is achieved by 
taking the derivative to $x$ and searching the singularities in
\begin{eqnarray}
h'(K,x)&=& \sum_{k=1}^{\infty} 2 (-1)^{k}  J_{k}(k K)
\cos(2 \pi k x)
\label{serdh}
\end{eqnarray}
instead of $h(K,x)$. This is allowed as the two problems are equivalent.

From the series~(\ref{serdh}) we can guess
for which values of $x$ the singularities will be $K_c$ and
$\rho$ respectively. 
As $J_k(kK)$ is positive for real values $0 < K < 1$ 
(see p. 534 in \cite{Watson1944},
we need to compensate the $(-1)^k$ term by $\cos(2\pi k x)$.
This is achieved  
for $x=\frac{1}{2}$ that reduces Eq.~(\ref{serdh}) to 
\begin{eqnarray}
h'(K,\frac{1}{2})&=& 2 \sum_{k=1}^{\infty}   J_{k}(k K) ,
\end{eqnarray}
which has the exact solution
(see formula (1) at p. 615 in \cite{Watson1944})
\begin{eqnarray}
2 \sum_{k=1}^{\infty}   J_{k}(k K) =\frac{K}{1-K} .
\end{eqnarray}
Hence, $h'(K,\frac{1}{2})$ has a singularity at $K=1$, yielding the
critical constant $K_c=1$, a well known result
in celestial mechanics~\cite{Wintner1941}.

\begin{figure}[!h]
\includegraphics[width=8cm,keepaspectratio]{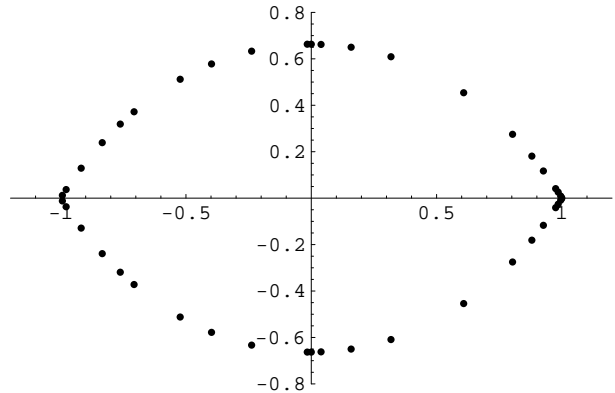}
\caption{Singularities in $K$ of the function $h(x,K)$ 
for the SM without small divisors for $x$ varying in $[0,1]$.
The radius of convergence $\rho$
corresponds to the value $x=1/4$, while the critical
constant $K_{c}=1$ corresponds to $x=1/2$. The curve is symmetric
with respect to both the real and the imaginary axes.}
\label{fig4}
\end{figure}

The complete analyticity domain can be found
in Ref.~\cite{Wintner1941}, p. 219. In Fig.~\ref{fig4})
we represent what can be obtained by using
Pad\'{e} approximants for some values of $x$. What emerges is that
the function $h(x,K)$ has for each value of $x$ a pair
of complex singularities closest to the origin
symmetric with respect to the real axis.
For $x$ going from $0$ to $1/2$ such singularities move continuously
from $-1$ to $1$ along two (symmetric) curves which pass through
the points $\pm i\rho$ at $x=1/4$ (see Fig.~\ref{fig4}).
Hence the entire set of singularities closest to
the origin lies on a curve which is smooth except at $K=\pm 1$,
where it has a discontinuity in its first derivative
(cf. again Ref.~\cite{Wintner1941}, p. 219).
An important feature is, however, that, as already noted in a similar
context by Simon~\cite{Simon1985}, a natural boundary
in $K$ for fixed $x$ seems to appear only in the
presence of small divisors. In fact, the latter give rise to
the occurrence of sudden peaks yielding a pattern similar to 
lacunary series~\cite{Katznelson1968}
for which natural boundaries can be proved to arise.
Hence, these peaks seem to be responsible of the formation
of the natural boundary as suggested by Prange
(cf. again Ref.~\cite{Simon1985}).

\section{Conclusions}
\label{Concl}
We showed by a numerical evaluation of the Lindstedt series
up to order $n=700$ that a previously assumed violation of Greene's criterion 
was ungrounded. The assumption
that allowed the restricted series~(\ref{recSSM})
was falsified for orders $n>200$.
The resulting critical constant did not correspond to the SM,
but is still true for the SSM.  
From our numerics, we conclude that, for the golden mean, the 
SSM critical constant is strictly higher than the SM.
This seems to be generally true for all winding numbers,
but is specifically difficult to proof for the golden mean
where both constants are very close.
Still, the numerics till order $n=700$ 
do not give a complete convergence.  
An evaluation that would compare to the accuracy of Greene's method
would rely on an prohibitive computational effort.

In addition to this analysis, we have purposed a new model 
that appears when the small denominators in the SM and SSM
are suppressed. 
We show that this model maintains many features of the SM and SSM.
However, it has an analytical solution and corresponds to
Kepler's equation in case of the SM.
Also here, the analogue of the SM has a lower value than the one 
of the SSM. Moreover, surprisingly the radii of convergence are lower
than the true models for golden mean winding numbers.
This proofs that the golden mean winding numbers are remarkably resistant
to the small denominator effect and falsifies 
a common misconception that the small denominator problem
is the dominant and only mechanism for the breaking analyticity transition.
The fact, that the simplified model still has a transition with a value
even lower than the true maps for the golden mean, shows that this is
not the case.

Finally, we studied the full analyticity domain for the two models.
Also here, there are striking differences between the simplified- and
the true maps. Similar to the true maps, the set of singularities
form a natural boundary. However, whereas the SM and SSM all the singularities
in $K$ for the function  $g(K,x)$ are present for any value of $x$,
the situation is different for the simplified maps.  
The simplified SSM has only one singularity in the complex $K$-plane 
for the function $h(K,x)$ at each value of $x$. 
The simplified SM has for each value of $x$ two singularities
symmetric with respect to the real axis, except for the singularities
on the real axis for $x=0$ and $x=1/2$. The closed natural boundary
is retained after gathering all singularities for all $x$. 

This natural boundary is perfect circle in case of the simplified 
SSM, while it is more stretched curve for the simplified SM
with a discontinuity on the real axis at $K=\pm 1$.
This shows that the radius of convergence of the simplified SSM equals
its critical constant, like was found for the true maps with
golden mean winding numbers. In contrast, the simplified SM has 
a critical constant of $K_c=1$ that is higher than its
radius of convergence. In that respect, the simplified SM resembles
more the true SM with winding numbers close to rational values.
Also this is a bit of a surprise, as one would expect the contrary,
but is consistent with the trend mentioned above.
Somehow, it is almost as if
the model, in which all small divisors were eliminated, still suffers more
from this effect that the SM with the golden mean.

Therefore, we believe that the study of these kind of simplified 
analytical models are a worthy prerequisite for the understanding
of the SM, SSM and FK models and, in particular, the influence of the
small denominators.

\acknowledgements
We thank Alberto Berretti for pointing out the relation
between the simplified SM and Kepler's equation. T.S.v.E acknowledges
the support by a Marie Curie Intra-European Fellowships
(MEIF-CT-2003-501976) within
the 6th European Community Framework Programme and the support through
the European Network LOCNET HPRN-CT-1999-00163.

\appendix
\section{induction proof}
\subsection{Proof of Eq.~(\ref{relbinomial})}
\label{secindR}
Assuming that relation (\ref{relbinomial}) is true
up to some Taylor order $n-1$, then the first relation of 
Eq.~(\ref{recR}) for $n$ yields:
\begin{eqnarray}
R(n,1) &=& n \sum_{m=1}^{n-1} \frac{(n-1)^{n-1-m}}{(m-1)!}
\frac{(n-2)!}{(n-1-m)!} \nonumber \\
&=& n \sum_{m=0}^{n-2} \frac{(n-1)^{m}}{(n-2-m)!}
\frac{(n-2)!}{m!} \nonumber \\
&=& n \sum_{m=0}^{n-2} (n-1)^m \binom{n-2}{m} \nonumber \\
&=&  n \big( (n-1) + 1 \big)^{n-2}=n^{n-1} ,
\end{eqnarray}
where in third equality we have used the binomial theorem 
\begin{eqnarray}
(1+x)^n =\sum_{j=0}^n \binom{ n}{j} x^j .
\end{eqnarray}
The second relation of Eq.~(\ref{recR}) is more difficult.
One can write
\begin{eqnarray}
R(n,m) &=&
\frac{1}{m} \sum_{n'=1}^{n-m+1} \binom{n}{n'}\binom{n-n'-1}{m-2}
\nonumber \\
&\times&
n'^{(n'-1)} (n-n')^{(n-n'-m+1)}
\nonumber \\
&=&\frac{1}{m} \sum_{n'=1}^{n-m+1} \frac{n!}{n'!(n-n')!}
\frac{(n-n'-1)!}{(m-2)!(n-n'-m+1)!} 
\nonumber \\
&\times&
n'^{(n'-1)} (n-n')^{(n-n'-m+1)}
\nonumber \\
&=&\frac{1}{m} \sum_{n'=1}^{n-m+1}  
\frac{n!}{n'!(m-2)!(n-n'-m+1)!}  
\nonumber \\
&\times&
n'^{(n'-1)}
(n-n')^{(n-n'-m)} \nonumber \\
&=&
\frac{1}{m} \sum_{n'=1}^{n-m+1}  \frac{n!}{ (m-2)!(n-m+1)!}
\binom{n-m+1}{n'}
\nonumber \\
&\times&
n'^{(n'-1)} (n-n')^{(n-n'-m)} \nonumber \\
&=& \frac{m-1}{m} \sum_{n'=1}^{n-m+1}  \binom{n}{m-1} \binom{n-m+1}{n'}
\nonumber \\
&\times&
n'^{(n'-1)} (n-n')^{(n-n'-m)} \nonumber \\
&=& \frac{m-1}{m} \binom{n}{m-1} \sum_{n'=0}^{n-m}  \binom{n-m+1}{n'+1}
\nonumber \\
&\times&
(n'+1)^{n'} (n-n'-1)^{(n-n'-1-m)} \nonumber \\
&=&
\frac{(m-1)(n-m+1)}{m} \binom{n}{m-1} \sum_{n'=0}^{n-m}  \binom{n-m}{n'}
\nonumber \\
&\times&
(n'+1)^{n'-1} (n-n'-1)^{(n-n'-1-m)} .
\end{eqnarray}
Using Abel's identity~\cite{Riordan79}
\begin{eqnarray}
(x+y)(x+y-a{\tilde n})^{{\tilde n}-1}&=&\sum_{k=0}^{\tilde n}
\binom{\tilde n}{k} xy (x-ak)^{k-1} \nonumber \\
&\times&
[y-a({\tilde n}-k)]^{{\tilde n}-k-1} ,
\end{eqnarray}
with $k=n', {\tilde n}=n-m, a=-1, x=1, y=m-1$ yields
\begin{eqnarray}
m n^{n-m-1}&=&\sum_{n'=0}^{n-m}
\binom{n-m}{n'} (m-1)  \\ &\times&
(1+n')^{n'-1}
(n-n'-1)^{n-m-n'-1}. \nonumber
\end{eqnarray}
Hence
\begin{eqnarray}
R(n,m)&=&
\frac{(m-1)(n-m+1)}{m} \binom{n}{m-1} \sum_{n'=0}^{n-m}  \binom{n-m}{n'}
\nonumber \\ &\times&
(n'+1)^{n'-1} (n-n'-1)^{(n-n'-1-m)} \nonumber \\
&=& (n-m+1) \binom{n}{m-1}  n^{n-m-1}
\nonumber \\
&=&
\frac{n-m+1}{n} \frac{n!}{(m-1)!(n-m+1)!}  n^{n-m} \nonumber \\
&=&\frac{(n-1)!}{(m-1)!(n-m)!}  n^{n-m}=\binom{n-1}{m-1} n^{n-m} , \nonumber \\
\end{eqnarray}

\section{tree formalism}
\subsection{Proof of Eq.~(\ref{relbinomial})}
\label{sectreeR}
First of all, by defining $\alpha = 2\pi x$
and $u(\alpha)=2\pi g(x)$, one can write the functional relation
that the function $u(\alpha)$ has to satisfy as
$u(\alpha) + (K/2i) \exp (i\alpha+iu(\alpha))=0$ for the SSM and
$u(\alpha) + K \sin(\alpha+u(\alpha))=0$ for the SM.
Note that in the case of the SSM the function $u$, which in principle
depends on two parameters $K$ and $\alpha$, is in fact a
function of the only parameter $\eta\equiv Ke^{i\alpha}$.

In terms of the function $u(\alpha)$ the functional equation
(\ref{functionalequation}) becomes, for the SM,
\begin{equation}
2u(\alpha)-u(\alpha+2\pi l)-u(\alpha-2\pi l) = - K \sin
(\alpha+u(\alpha)) ,
\label{functionalequationu}
\end{equation}
in which we recognize Eq. (1.4) of Ref. \cite{BerGen1999},
with $\varepsilon=K$ and $\omega=l$. For the SSM we have the
same equation with the sine function
replaced with $(2i)^{-1}\exp(i\alpha+iu(\alpha))$.
Then we can envisage the same tree expansion as in Ref. \cite{BerGen1999};
see formula (2.2), where, to make a relation
with the notations we are using now, $k$ and $\nu$
are what we are denoting with $n$ and $k$, respectively.
Moreover $\gamma(nu_{\ell_{v}})=-\omega_{\nu_{\ell_{v}}}$,
hence it is $-1$ in our case, and one has $\nu_{v}=1$ for the SSM
and $\nu_{u}\in\{\pm1\}$ for the SM. At the end we find
\begin{equation}
X^{n}_{k} = \frac{1}{2\pi i} \frac{(-1)^{n}}{2^{n}}
\sum_{\vartheta \in \mathcal{T}_{n,k}}
\operatorname{Val} (\vartheta), \qquad
\operatorname{Val} (\vartheta) =
 \prod_{u \in \vartheta} \frac{1}{m_{u}!} \nu_{u}^{m_{u}+1} ,
\label{smtree}
\end{equation}
where the trees $\vartheta$, the branching numbers $m_{u}$ and the set of
trees $\mathcal{T}_{n,k}$ of order $n$
(that is with $n$ nodes) and with momentum $k$
flowing through the root line (that is such that
$\sum_{u\in\vartheta}\nu_{u}=k$) are defined as in Ref. \cite{BerGen1999}.

In the case of the SSM, Eq. (\ref{smtree}) reduces to
\begin{equation}
X^{n}_{n} = \frac{1}{2\pi i} \frac{(-1)^{n}}{2^{n}}
\sum_{\vartheta\in \mathcal{T}_{n,n}}
\operatorname{Val}(\vartheta), \qquad
\operatorname{Val}(\vartheta) = \prod_{u \in \vartheta} \frac{1}{m_{u}!} ,
\label{ssmtree}
\end{equation}
as $\nu_{u}\equiv 1$, and the sum over trees of order $n$
can be written as a sum over all possible configurations of
branching numbers $\{m_{u}\}_{u\in \vartheta}$
with the constraint $\sum_{u\in \vartheta} m_{u}=n-1$:
indeed they are the only labels of the trees, and their values
uniquely determine the elements of $\mathcal{T}_{n,n}$.
Therefore we can rewrite $X^{n}_{n}$ as
\begin{eqnarray}
X^{n}_{n} &=& \frac{1}{2\pi i} \frac{(-1)^{n}}{2^{n}}
\sum_{m_{1}+\ldots+m_{n}=n-1}
\frac{1}{m_{1}!\ldots m_{n}!}  \nonumber \\
&=& \frac{1}{2\pi i}
\frac{(-1)^{n}}{2^{n}} \frac{n^{n-1}}{n!} ,
\label{ssmcoeff}
\end{eqnarray}
where we have used the multinomial theorem
\begin{equation}
\sum_{m_{1}+\ldots+m_{n}=p} \frac{n!}{m_{1}!\ldots m_{n}!}
x_{1}^{m_{1}} \ldots x_{n}^{m_{n}} = \left( x_{1}+\ldots+x_{n}
\right)^{p} ,
\label{multinomial}
\end{equation}
which extends the binomial theorem to $n>2$; see \cite{AbrSte1972},
\S 24.1.3.

\subsection{Proof of Eq.~(\ref{resultPnk1})}
\label{treeproof2}
In the case of the SM, without small divisors, we can still
use formula (\ref{smtree}),
but now one can have $\nu_{u}=\pm 1$.
\begin{eqnarray}
X^{n}_{k} & = & \frac{1}{2\pi i} \frac{(-1)^{n+(n-k)/2}}{2^{n}}
\binom{n}{(n-k)/2} \nonumber \\
&&\sum_{m_{1}+\ldots+m_{n}=n-1}
\frac{\nu_{1}^{m_{1}} \ldots \nu_{n}^{m_{n}}}{m_{1}!\ldots m_{n}!}
\\
& = & \frac{1}{2\pi i} \frac{(-1)^{n+(n-k)/2}}{2^{n}}
\frac{k^{n-1}}{((n-k)/2)!((n+k)/2)!} , \nonumber
\label{smcoeff}
\end{eqnarray}
for $k=n-2p$, with $p=1,\ldots,n$.

By using the definition Eq. (\ref{rocdavie}) of radius
of convergence, we see that we have to compute
the maximum over $k$ of $|X^{n}_{k}|$.

As $k=\sum_{u\in\vartheta}\nu_{u}$ we see that, first, $k$ can assume only
the values $-n,-n+2,-n+4,\ldots,n-4,n-2,n$ (so that, in particular,
$(n\pm k)/2$ is even), and, second, in order to have a contribution to
$X^{n}_{k}$ we have to put $(n-k)/2$ mode labels $\nu_{u}$ equal to
$-1$ and the remaining $(n+k)/2$ mode labels equal to $1$.
Moreover for any tree $\vartheta\in\mathcal{T}_{n,k}$ we can write
\begin{equation}
\prod_{u \in \vartheta} \nu_{u}^{m_{u}+1} =
\Big( \prod_{u \in \vartheta} \nu_{u} \Big)
\Big( \prod_{u \in \vartheta} \nu_{u}^{m_{u}} \Big) =
(-1)^{(n-k)/2} \prod_{u \in \vartheta} \nu_{u}^{m_{u}} ,
\label{numerator}
\end{equation}
which inserted into Eq. (\ref{smtree}) gives,
by using again the multinomial theorem,
\begin{eqnarray}
X^{n}_{k} & = & \frac{1}{2\pi i} \frac{(-1)^{n+(n-k)/2}}{2^{n}}
\binom{n}{(n-k)/2} \nonumber \\
&&\sum_{m_{1}+\ldots+m_{n}=n-1}
\frac{\nu_{1}^{m_{1}} \ldots \nu_{n}^{m_{n}}}{m_{1}!\ldots m_{n}!}
\\
& = & \frac{1}{2\pi i} \frac{(-1)^{n+(n-k)/2}}{2^{n}}
\frac{k^{n-1}}{((n-k)/2)!((n+k)/2)!} , \nonumber
\label{smcoeffbis}
\end{eqnarray}

\bibliographystyle{prsty}

\begin{thebibliography}{10}

\bibitem{Chi79}
B.~V. Chirikov, Phys. Rep. {\bf 52},  263  (1979).

\bibitem{Gre1979}
J.~M. Greene, J. Math. Phys. {\bf 20},  1183  (1979).

\bibitem{FloMa96}
L.~M. Floria and J.~J. Mazo, Adv. Phys. {\bf 45},  505  (1996).

\bibitem{Tabor89}
M. Tabor, {\em Chaos and Integrability in Nonlinear Dynamics: An introduction}
  (Wiley, New York, 1989).

\bibitem{Math84}
J.~N. Mather, Ergod. Theor. and Dynam. Sys. {\bf 4},  301  (1984).

\bibitem{MaPer85}
R.~S. MacKay and I.~C. Percival, Commun. Math. Phys. {\bf 98},  469  (1985).

\bibitem{Jung91}
I. Jungreis, Ergod. Theor. and Dynam. Sys. {\bf 11},  79  (1991).

\bibitem{MacK82}
R.~S. MacKay, {\em Renormalisation in area preserving maps}, {\em Advanced
  Series in Nonlinear Dynamics} (World Scientific, Singapore, 1993).

\bibitem{Faldela1992}
C. Falcolini and R. de~la Llave, J. Stat. Phys. {\bf 67},  609  (1992).

\bibitem{Deldela2000}
A. Delshams and R. de~la Llave, SIAM. J. Math. Anal. {\bf 31},  1235  (2000).

\bibitem{ErpFas2002}
T.~S. van Erp and A. Fasolino, Europhys. Lett. {\bf 59},  330  (2002).

\bibitem{Aubry83}
S. Aubry, Physica D {\bf 7},  240  (1983).

\bibitem{ErpFas99}
T.~S. van Erp, A. Fasolino, O. Radulescu, and T. Janssen, Phys. Rev. B {\bf
  60},  6522  (1999).

\bibitem{BerCelChiFal1992}
A. Berretti, A. Celletti, L. Chierchia, and C. Falcolini, J. Stat. Phys. {\bf
  66},  1613  (1992).

\bibitem{BerFalGen2001}
A. Berretti, C. Falcolini, and G. Gentile, Phys. Rev. E {\bf 64},  015202
  (2001).

\bibitem{Faldela1992b}
C. Falcolini and R. de~la Llave, J. Stat. Phys. {\bf 67},  645  (1992).

\bibitem{Wintner1941}
A. Wintner, {\em The analytic foundations of celestial mechanics} (Princeton
  University Press, Princeton, NJ, 1941).

\bibitem{Siegel1942}
C.~L. Siegel, Ann. of Math. {\bf 43},  607  (1942).

\bibitem{Dav1995}
A.~M. Davie, Nonlinearity {\bf 7},  219  (1994).

\bibitem{BerGen1999}
A. Berretti and G. Gentile, J. Math. Pure Appl. {\bf 78},  159  (1999).

\bibitem{BerGen2001}
A. Berretti and G. Gentile, Comm. Math. Phys. {\bf 220},  623  (2001).

\bibitem{WeiEl96}
M. Weiss and F.~J. Elmer, Phys. Rev. B {\bf 53},  7539  (1996).

\bibitem{WeiEl97}
M. Weiss and F.~J. Elmer, Z. Phys. B: Condens. Matter {\bf 104},  55  (1997).

\bibitem{GyaTho97}
T. Gyalog and H. Thomas, Eur. Phys. Lett. {\bf 37},  195  (1997).

\bibitem{Gal1994}
G. Gallavotti, Comm. Math. Phys. {\bf 164},  145  (1994).

\bibitem{Wells86}
D. Wells, {\em The Penguin Dictionary of Curious and Interesting Numbers}
  (Penguin Books, Middlesex, England, 1986).

\bibitem{WW1921}
E.~T. Whittaker and G.~N. Watson, {\em A course of modern analysis} (Cambridge
  University Press, Cambridge, 1997).

\bibitem{Finch2003}
S.~R. Finch, {\em Mathematical constants} (Cambridge University Press,
  Cambridge, 2003).

\bibitem{Wilbrink1987}
J. Wilbrink, Physica D {\bf 26},  358  (1987).

\bibitem{GraRyz2000}
I.~S. Gradshteyn and I.~M. Ryzhik, {\em Table of integrals, series, and
  products} (Academic Press, San Diego, 2000).

\bibitem{Watson1944}
G.~N. Watson, {\em A treatise on the theory of Bessel functions} (Cambridge
  University Press, Cambridge, 1944).

\bibitem{Simon1985}
B. Simon, Ann. of Phys. {\bf 159},  157  (1985).

\bibitem{Katznelson1968}
Y. Katznelson, {\em An introduction to harmonic analysis}, {\em Cambridge
  Mathematical Library} (Cambridge University Press, Cambridge, 2004).

\bibitem{Riordan79}
J. Riordan, {\em Combinatorial Identities} (Wiley, New York, 1979).

\bibitem{AbrSte1972}
A.~M. Abramowitz and I.~A. Stegun, {\em Handbook of Mathematical functions}
  (Dover Publications, New York, 1972).

\bibitem{Poin1899}
H. Poincar\'e, {\em Les m\'ethodes nouvelles de la m'ecanique classique, Vol.
  III} (Gauthier-Villars, Paris, 1899).

\bibitem{Tom29}
G.~A. Tomlinson, Philos. Mag. {\bf 7},  905  (1929).

\end{thebibliography}

\end{document}